\documentclass[creativecommons]{eptcs/eptcs}
\providecommand{\event}{LFMTP 2015}

\usepackage[pdftex]{virginialake}
\usepackage{mathdesign}
\usepackage{amsthm}
\usepackage{graphics}
\usepackage{microtype}
\usepackage{eucal}
\usepackage{eufrak}
\usepackage{xspace}

\newcommand{\rnm}[1]{\textbf{\textit{#1}}}

\newcommand{\cmtout}[1]{}

\newcommand{\qqquad}[0]{\quad\quad\quad}

\newcommand{\grdef}[0]{\mathrel{\raise.5pt\hbox{$\mathop{::}$}{=}}}
\newcommand{\mmid}{\:\mid\:}

\newcommand{\gseq}[2]{{#1 \hspace*{1pt}\vdash #2}}
\newcommand{\fseq}[2]{{#1 \hspace*{1pt}\vDash #2}}

\newcommand{\msub}[1]{\{#1\}}
\newcommand{\rdc}[0]{\mathbin{\,\rightarrow\,}}

\newcommand{\pr}[1]{\langle\hspace*{0.1ex}#1\hspace*{0.1ex}\rangle}

\newcommand{\don}[0]{\mathord{\triangleright\;}}
\newcommand{\nod}[0]{\mathord{\triangleleft\;}}
\newcommand{\din}[0]{\mathord{\scalebox{0.9}{\texttt{\,in\,}}}\xspace}
\newcommand{\pcp}[0]{\mathord{\scalebox{0.9}{\texttt{\,@\,}}}\xspace}
\newcommand{\typ}[0]{\scalebox{0.9}{\texttt{type}}}
\newcommand{\els}[0]{\varepsilon}
\newcommand{\inl}[0]{\mathord{\scalebox{0.9}{\texttt{inl\,\,}}}\xspace}
\newcommand{\inr}[0]{\mathord{\scalebox{0.9}{\texttt{inr\,\,}}}\xspace}
\newcommand{\prl}[0]{\mathord{\scalebox{0.9}{\texttt{prl\,\,}}}\xspace}
\newcommand{\prr}[0]{\mathord{\scalebox{0.9}{\texttt{prr\,\,}}}\xspace}

\newcommand{\spt}[3]{#1[#2 \mid #3]}

\newcommand{\trm}[1]{{\color{red} #1}}

\newcommand{\LJ}{\textbf{{LJ}}\xspace}
\newcommand{\LJT}{\textbf{{LJT}}\xspace}
\newcommand{\LJF}{\textbf{{LJF}}\xspace}
\newcommand{\muLJ}{\textbf{{$\mu$LJ}}\xspace}
\newcommand{\muMALL}{\textbf{{$\mu$MALL}}\xspace}
\newcommand{\lbar}[0]{\smash{\overline{\lambda}}}

\newcommand{\imp}[0]{\rightarrow\xspace}
\newcommand{\nfy}[0]{\mathord{\uparrow}}
\newcommand{\pfy}[0]{\mathord{\downarrow}}

\newcommand{\iaxrule}[2]{
  \vlinf{\rnm{#1}}{}{#2}{}}
\newcommand{\irule}[3]{
  \vlinf{\rnm{#1}}{}{#2}{#3}}

\newcommand{\iruule}[4]{
  \vliinf{\rnm{#1}}{}{#2}{#3}{#4}}

\newcommand{\iruuuule}[6]{
  \vliiinf{\rnm{#1}}{}{#2}{#3}{#4}{#5 \quad #6}}

\newcommand{\gseqrule}[5]{
  \irule{#1}{\gseq{#2}{#3}}{\gseq{#4}{#5}}}

\newcommand{\gseqruule}[7]{
  \iruule{#1}{\gseq{#2}{#3}}{\gseq{#4}{#5}}{\gseq{#6}{#7}}}

\newcommand{\fseqaxrule}[3]{
  \iaxrule{#1}{\fseq{#2}{#3}}}
\newcommand{\fseqrule}[5]{
  \irule{#1}{\fseq{#2}{#3}}{\fseq{#4}{#5}}}

\newcommand{\fseqruule}[7]{
  \iruule{#1}{\fseq{#2}{#3}}{\fseq{#4}{#5}}{\fseq{#6}{#7}}}

\title{Sequent Calculus and Equational Programming \\
  {\rm \normalsize (work in progress)}}
\def\titlerunning{Sequent Calculus and Equational Programming}
\author{Nicolas Guenot and Daniel Gustafsson
\institute{IT University of Copenhagen}
\email{\{ngue,dagu\}@itu.dk}}
\def\authorrunning{N.Guenot and D. Gustafsson}

\begin{document}
\vlnosmallleftlabels

\maketitle

\begin{abstract}
Proof assistants and programming languages based on type theories usually come
in two flavours: one is based on the standard natural deduction presentation
of type theory and involves eliminators, while the other provides a syntax in
\emph{equational} style. We show here that the equational approach corresponds
to the use of a focused presentation of a type theory expressed as a sequent
calculus.~A typed functional language is presented, based on a sequent calculus,
that we relate to the syntax and internal language of Agda. In particular, we
discuss the use of patterns and case splittings, as well as rules implementing
inductive reasoning and dependent products and sums.
\end{abstract}

%%%%%%%%%%%%%%%%%%%%%%%%%%%%%%%%%%%%
%%%%%%%%%%% INTRODUCTION %%%%%%%%%%%
%%%%%%%%%%%%%%%%%%%%%%%%%%%%%%%%%%%%

\section{Programming with Equations}
\label{sec:progeq}

Functional programming has proved extremely useful in making the task of writing
correct software more abstract and thus less tied to the specific, and complex,
architecture of modern computers. This, is in a large part, due to its extensive
use of types as an abstraction mechanism, specifying in a crisp way the intended
behaviour of a program, but it also relies on its \emph{declarative} style, as
a mathematical approach to functions and data structures. However, the vast gain
in expressivity obtained through the development of \emph{dependent types} makes
the programming task more challenging, as it amounts to the question of proving
complex theorems --- as illustrated by the double nature of proof
assistants such as Coq~\cite{dowek:al:93:coq} and Agda~\cite{norell:phd}.
Keeping this task as simple as possible is then of the highest importance, and
it requires the use of a clear declarative style.

There are two main avenues for specifying a language of proofs, or programs,
that is abstract enough to support complex developments involving dependent
types. The first approach, chosen by the Coq project, is to have a language
of \emph{tactics} that partially automate the construction of proofs --- that
is, to mechanically construct complex programs based on the composition of a
few generic commands. While this takes the development task closer to the
usual idea of proving a mathematical theorem, the second approach is to take
the programming viewpoint: although Coq allows to directly write proof terms,
this is better illustrated by Agda, where a syntax inspired by Haskell
\cite{haskell:www} provides a clear \emph{equational} style.

Our goal here is to investigate the relations between the equational style
of dependently-typed functional programming as found in Agda to the
proof-theoretical description of intuitionistic logic given in the sequent
calculus. In particular, we claim that a \emph{focused} sequent calculus,
akin to the \LJF system of Liang and Miller
\cite{liang:miller:09:focpol}, offers a logical foundation of choice for the
development of a practical dependently-typed language. We intend to support this
claim by showing how the equational syntax of Agda and the internal structure
of its implementation correspond to a computational interpretation of such
a calculus --- for an extended for of intuitionistic logic including
dependencies and (co)induction. As it turns out, the use of left rules
rather than eliminations for \emph{positive} connectives such as disjunction,
in sequent calculus, yields a simpler syntax. In general, beyond the use
of \emph{spines} in applications, as in \LJT \cite{herbelin:94:chseq} and
quite common in the implementation of functional programming languages or
proof assistants, the structure of the sequent calculus is much closer to
the equational style of programming than natural deduction, the standard
formalism in which type theory is usually expressed \cite{martinloef:84:itt}.
Using~a focused system rather than a plain sequent calculus based on \LJ
provides a stronger structure, and emphasizes the importance of
\emph{polarities}, already observed in type theory
\cite{abel:pientka:thibodeau:setzer:14:copat}.

Beyond the definition of a logical foundation for a functional language in
equational style, giving a proof-theoretical explanation for the way Agda
is implemented requires to accomodate in the sequent calculus both dependent
types and a notion of inductive definition. This is not an easy task, although
there has been some work on dependent types in the sequent calculus
\cite{dyckhoff:lengrand:mckinna:11:focpts} and there is a number of
approaches to inductive definitions in proof theory, including focused
systems \cite{baelde:12:mumall}. For example, the system found in
\cite{dyckhoff:lengrand:mckinna:11:focpts} is based on \LJT but is limited
to $\Pi$ and does not support $\Sigma$, while \cite{dyckhoff:pinto:98:seqdep}
has both, but requires an intricate mixture of natural deduction and sequent
calculus to handle $\Sigma$. Induction is even more complex
to handle, since there are several approaches, including
definitions \cite{schroeder-heister:93:defr} or direct least and
greatest fixpoints as found in \muMALL \cite{baelde:12:mumall} and
\muLJ \cite{baelde:phd}. From the viewpoint of proof-theory, the least
fixpoint operator $\mu$ seems to be well-suited, as it embodies
the essence of induction, while the greatest fixpoint $\nu$ allows to
represent coinduction. However, these operators are not used the same way
as inductive definitions found in Agda or other languages or proof assistants
--- they seem more primitive, but the encoding of usual constructs in terms
of fixpoints is not obvious. Even more complicated is the question of using
fixpoints in the presence of dependent types, and this has only been studied
from the type-theoretic viewpoint in complex systems such as the \emph{Calculus
of Inductive Constructions} \cite{coquand:paulin:88:cic}. In the end, what
we would like to obtain is a proof-theoretical understanding of the
equational style of dependent and (co)inductive programming, related to the
goals of the Epigram project. In particular, we consider that the sequent
calculus, with its use of left rules, provides access to the
\emph{``left''} of equations in a sense similar to what is
described in \cite{mcbride:mckinna:04:left}.

Here, we will describe the foundamental ideas for using a variant of \LJF
as the basis for the design of a dependently-typed programming language.
We start in Section \ref{sec:focpsc} by considering a propositional system
and show how the shape of sequent calculus rules allows to type terms in
equational style.~This is made even more obvious by the use of pattern
in the binding structure of the calculus. Then, in Section \ref{sec:depind}
we discuss the extension of this system to support dependent types and
induction, problems related to patterns in this setting, as well as the
question of which proof-theoretical approach to induction and coinduction
is better suited for use in a such a language. Finally, we conclude by
the review of some research problems opened by this investigation, and
an evaluation of the possible practical applications to languages and
proofs assistants.

%%%%%%%%%%%%%%%%%%%%%%%%%%%%%%%%%%%%
%%%%%%%%%%%% FOCUSED SC %%%%%%%%%%%%
%%%%%%%%%%%%%%%%%%%%%%%%%%%%%%%%%%%%

\section{Focusing and Polarities in the Sequent Calculus}
\label{sec:focpsc}

We start our investigation with a propositional intuitionistic system presented
as a focused sequent calculus. It is a variant of \LJF \cite{liang:miller:09:focpol}
to which we assign a term language extending the $\lbar$-calculus of
Herbelin~\cite{herbelin:94:chseq}. Unlike the calculus based on \LJT, this system
has positive disjunctions and conjunctions $\lor$ and $\times$, but it has
no positive atoms. We use the following grammar of formulas:
$$N,M ~\grdef~ a \,\mmid\, \nfy P \,\mmid\, P \imp N \,\mmid\, N \land M
  \qquad\qquad
  P\,\!,Q ~\grdef~ \pfy N \,\mmid\, P \lor Q \,\mmid\, P \times Q$$
where $\nfy$ and $\pfy$ are called \emph{polarity shifts} and are meant to
maintain an explicit distinction between the two categories of formulas,
negatives and positives. This is not absolutely necessary, but it clarifies
the definition of a focused system by linking the \emph{focus} and \emph{blur}
rules to actual connectives. Note that this was also used in the presentation
of a computational interpretation of the full \LJF system
\cite{brockn:guenot:gustafsson:15:ljfoc}.

\newpage

\begin{figure}[t]
\centerline{
$\begin{array}{|@{\quad}c@{\quad~}|}
  \hline \\
  \begin{array}{@{~}c@{~}}
    \fseqaxrule{}{\Psi,[N]}{\trm{\els}:N} \qqquad
    \irule{}{\gseq{\Psi \mmid \cdot}{\trm{\don d}:\nfy P}}
      {\fseq{\Psi}{\trm{d}:[P]}} \qqquad
    \irule{}{\fseq{\Psi}{\trm{\nod t}:[\pfy N]}}
      {\gseq{\Psi \mmid \cdot}{\trm{t}:N}} \\
    \\
    \irule{}{\gseq{\Psi,\trm{x}:\pfy N \mmid \cdot}{\trm{x\;k}:M}}
      {\fseq{\Psi,\trm{x}:\pfy N,[N]}{\trm{k}:M}} \qquad
    \irule{}{\gseq{\Psi \mmid \Gamma,\trm{x}:\pfy N}{\trm{t}:M}}
      {\gseq{\Psi,\trm{x}:\pfy N \mmid \Gamma}{\trm{t}:M}} \qquad
    \irule{}{\fseq{\Psi,[\nfy P]}{\trm{\kappa p.t}:N}}
      {\gseq{\Psi \mmid \trm{p}:P}{\trm{t}:N}} \\
    \\
    \begin{array}{c@{\qquad}c}
      \gseqrule{}{\Psi \mmid \Gamma}{\trm{\lambda p.\,t}:P \imp N}
        {\Psi \mmid \Gamma,\trm{p}:P}{\trm{t}:N} &
      \fseqrule{}{\Psi,[N \land\, M]}{\trm{\prl k}:L}
        {\Psi,[N]}{\trm{k}:L} \quad
      \fseqrule{}{\Psi,[N \land\, M]}{\trm{\prr k}:L}
        {\Psi,[M]}{\trm{k}:L} \\
      \\
      \iruule{}{\fseq{\Psi,[P \imp N]}{\trm{d::k}:M}}
        {\fseq{\Psi}{\trm{d}:[P]} \quad}{\fseq{\Psi,[N]}{\trm{k}:M}} &
      \gseqruule{}{\Psi \mmid \Gamma}{\trm{\pr{t,u\,}}:N \land\, M}
        {\Psi \mmid \Gamma}{\trm{t}:N \quad}
        {\Psi \mmid \Gamma}{\trm{u}:M} \\
    \end{array} \\
    \\
    \begin{array}{c@{\qquad}c}
      \fseqruule{}{\Gamma}{\trm{(d,e)}:[P \times\, Q]}
        {\Gamma}{\trm{d}:[P] \quad}{\Gamma}{\trm{e}:[Q]} &
      \fseqrule{}{\Gamma}{\trm{\inl d}:[P \,\lor\, Q]}
        {\Gamma}{\trm{d}:[P]} \quad
      \fseqrule{}{\Gamma}{\trm{\inr d}:[P \,\lor\, Q]}
        {\Gamma}{\trm{d}:[Q]} \\
      \\
      \gseqrule{}{\Psi \mmid \Gamma,\trm{(p,q)}:P \times\, Q}{\trm{t}:N}
        {\Psi \mmid \Gamma,\trm{p}:P,\trm{q}:Q}{\trm{t}:N} &
      \gseqruule{}{\Psi \mmid \Gamma,\trm{\spt{x}{p\!}{q}}:P \,\lor\, Q}
        {\trm{\spt{x}{t}{u}}:N}
        {\Psi \mmid \Gamma,\trm{p}:P}{\trm{t}:N \quad}
        {\Psi \mmid \Gamma,\trm{q}:Q}{\trm{u}:N} \\
    \end{array} \\
  \end{array} \\
  \\
  \hline
\end{array}$}
\caption{Typing rules for a pattern-based $\lambda$-calculus based on $\lbar$}
\label{figproplam}
\end{figure}

The rules we use in this system are shown in Figure \ref{figproplam}, where
the term assignment is indicated in red and several turnstiles are used to
distinguish an inversion phase $\vdash$ from a focused phase $\vDash$. In this
syntax, brackets are used to pinpoint the precise formula under focus. The
extended $\lambda$-calculus we use to represent proofs is based on the
following grammar:
$$\begin{array}{r@{~\grdef~}c@{~\mmid~}c@{~\mmid~}l}
  t,u & \don d & \lambda p.t & x\;k \,\mmid\, \pr{t,u}
    \,\mmid\, \spt{x}{t}{u} \\[0.1em]
  p,q & x & (p,q) & \spt{x}{p}{q} \\[0.1em]
  d,e & \nod t & (d,e) & \inl d \;\mmid\; \inr d \\[0.1em]
  k,m & \els & t::k & \hspace*{0.5pt}\prl k \;\mmid\; \prr k \,\mmid\, \kappa p.t \\
\end{array}$$
where $t$ denotes a \emph{term}, $p$ a binding \emph{pattern}, $d$ a \emph{data}
structure and $k$ an application \emph{context}. In terms of programming, terms are
describing computation, mostly by means of functions, while data structures
implement pairs and constructors. Note that computations can use \emph{case
splittings} $\spt{x}{t}{u}$ to choose between the subterms $t$ or $u$ depending
on the contents of the data bound to $x$. The use of patterns rather than plain
variables to annotate formulas in the context of typing judgement is taken
from \cite{cerrito:kesner:99:patcut} and allows to express more directly the
equational style found in Agda. For example, we could write:
$$\begin{array}{l}
  \mathtt{f~:~(\mathbb{N}\;\times\;\mathbb{N})~\uplus~\mathbb{N}~\imp~\mathbb{N}} \\
  \mathtt{f~(inl~(x,y))~=~x~+~y} \\
  \mathtt{f~(inr~z)~=~z}
\end{array}$$
to define a function $f$ that uses pattern-matching on its argument and
computes the result based on the components of the data structure it
received. Such a function can be written in our calculus as the following
term: $\lambda \spt{w}{(x,y)}{z}.
\spt{w}{\mathrm{add}\;((x\;\els)::(y\;\els)::\els)}{z\;\els}$,
where $\mathrm{add}$ is the name of the addition function. This makes
the compilation of the code written above to the adequate representation
in our calculus relatively easy, since different parts of a definition
can be aggregated into a term with a pattern and a case splitting. This
is very much related to the question of compiling pattern-matching into
a specific \emph{splitting tree} where case constructs are used
\cite{augustsson:85:compat}.

The idea of the logical approach is that \emph{cut elimination} in this
system yields a reduction system implementing the dynamics of computation
in the corresponding calculus. In such a focused calculus, a number of
cut rules are needed to complete the proof of completeness of the cut-free
fragment, but only two of them really need to be considered as rules ---
the other cuts can simply be stated as principles, and their reduction will
correspond to a big step of computation. These two rules are:
$$\iruule{}{\gseq{\Psi \mmid \Gamma}{\trm{p=d \din t}:N}}
    {\fseq{\Psi}{\trm{d}:[P]} \quad}
    {\gseq{\Psi \mmid \Gamma,\trm{p}:P}{\trm{t}:N}}
  \qquad\quad
  \iruule{}{\gseq{\Psi \mmid \Gamma}{\trm{t\;k}:M}}
    {\gseq{\Psi \mmid \Gamma}{\trm{t}:N} \quad}
    {\fseq{\Psi,[N]}{\trm{k}:M}}$$
the first one being the binding of a data structure to a matching
pattern, and the second a simple application of a term to a list of
arguments. The latter is already part of the \LJT system
\cite{herbelin:94:chseq}, but the former is specific to \LJF in the sense
that it appears only when formulas can be focused on the right of a sequent.
The main reduction rule extracted from cut elimination is the $\lbar$
variant of $\beta$-reduction:
$$(\lambda p.t)\;(d :: k) ~~\rdc~~ (p=d \din t)\;k$$
but there are a number of other reduction rules generated by the use of
other connectives than implication. In particular, conjunction yields a
form of pairing where a term $\pr{t,u}$ has to be applied to a list
$\prl k$ to reduction to $t\;k$. The binding cut is simpler in a certain
sense, since its reduction corresponds to a decomposition of the data
structure $d$ according to the shape of the pattern $p$, and a simple
substitution when $p$ is just a variable. Moreover, other cuts encountered
during reduction usually amount to a form of substitution, except for the
one, already present in \LJT, that yields lists concatenation in the argument
of an application.

Note that the $\don d$ construct is present in the internal language of Agda,
but the constructs $\nod t$ and $\kappa p.t$ are not, although they can be
obtained indirectly using a cut. While $\nod t$ should simply be understood
as a \emph{thunk}, which is a term made into data, the list $\kappa p.t$
is slightly more complex. This construct, already present in
\cite{barendregt:ghilezan:00:lamndseq}, is more a \emph{context} than a
list in the sense that it stops the application of a term to $\kappa p.t$
and enforces the execution of $t$, where the original term applied is bound
to $p$. This can be understood by considering the reduction extracted from
cut elimination:
$$(\don d)\;(\kappa p.t) ~~\rdc~~ p=d \din t$$

Finally, note that we could have an explicit contraction rule in the system,
that would appear in terms under the form of a pattern $p \!\pcp q$ indicating
that $p$ and $q$ will be the patterns associated to two copies of the same
assumption $P$. The associated typing rule is:
$$\gseqrule{}{\Psi \mmid \Gamma,\trm{p \!\pcp q}:P}{\trm{t}:N}
    {\Psi \mmid \Gamma,\trm{p}:P,\trm{q}:P}{\trm{t}:N}$$
and it is reminiscent of the pattern using the same syntax in Haskell ---
which is meant to exist in Agda as well, but this not yet implemented. However,
in Haskell, this is restricted to the form $x \pcp p$ so that it can only
serve to name an assumption before decomposing it, and we could allow for
such a use by avoiding maximal inversion, which is not strictly necessary
in a focused system \cite{brockn:guenot:gustafsson:15:ljfoc}. This rule is
not necessary for the completeness of the calculus, and there are other ways
to obtain the same result. Of course, in a very similar way, the pattern
\raisebox{0.1em}{$\texttt{\_}$} can be associated to the weakening rule,
also admissible.

\newpage

%%%%%%%%%%%%%%%%%%%%%%%%%%%%%%%%%%%%
%%%%%%%%%%%% DEP TYPES! %%%%%%%%%%%%
%%%%%%%%%%%%%%%%%%%%%%%%%%%%%%%%%%%%

\section{Adding Dependent Types and Induction}
\label{sec:depind}

\begin{figure}[t]
\centerline{
$\begin{array}{|@{\quad}c@{\quad~}|}
  \hline \\
  \begin{array}{@{}c@{}}
    \begin{array}{@{}c@{~\quad}c@{}}
      \fseqaxrule{}{\Psi,[N]}{\trm{\els}:N} \qqquad
      \irule{}{\gseq{\Psi \mmid \cdot}{\trm{\don d}:\nfy P}}
        {\fseq{\Psi}{\trm{d}:[P]}} \qqquad
      \irule{}{\fseq{\Psi}{\trm{\nod t}:[\pfy N]}}
        {\gseq{\Psi \mmid \cdot}{\trm{t}:N}} &
      \fseqrule{}{\Psi,[N \land\, M]}{\trm{\prl k}:L}
        {\Psi,[N]}{\trm{k}:L} \\
     \\
      \irule{}{\gseq{\Psi,\trm{x}:\pfy N \mmid \cdot}{\trm{x\;k}:M}}
        {\fseq{\Psi,\trm{x}:\pfy N,[N]}{\trm{k}:M}} \quad~
      \irule{}{\gseq{\Psi \mmid \Gamma,\trm{x}:\pfy N}{\trm{t}:M}}
        {\gseq{\Psi,\trm{x}:\pfy N \mmid \Gamma}{\trm{t}:M}} \quad~
      \irule{}{\fseq{\Psi,[\nfy P]}{\trm{\kappa x.t}:N}}
        {\gseq{\Psi \mmid \trm{x}:P}{\trm{t}:N}} &
      \fseqrule{}{\Psi,[N \land\, M]}{\trm{\prr k}:L}
        {\Psi,[M]}{\trm{k}:L} \\
    \end{array} \\
    \\
    \begin{array}{@{\!}c@{\!}}
      \gseqrule{}{\Psi \mmid \Gamma}{\trm{\lambda x.\,t}:\Pi(x:P). N}
        {\Psi \mmid \Gamma,\trm{x}:P}{\trm{t}:N} ~~
      \iruule{}{\fseq{\Psi,[\Pi(x:P).N]}{\trm{d::k}:M}}
        {\fseq{\Psi}{\trm{d}:[P]} \quad}{\fseq{\Psi,[N\msub{d/x}]}{\trm{k}:M}} ~~
      \gseqruule{}{\Psi \mmid \Gamma}{\trm{\pr{t,u\,}}:N \land\, M}
        {\Psi \mmid \Gamma}{\trm{t}:N \quad}
        {\Psi \mmid \Gamma}{\trm{u}:M} \\
    \end{array} \\
    \\
    \begin{array}{@{}c@{\quad}c@{}}
      \gseqrule{}{\Psi \mmid \Gamma,\trm{x}:\Sigma(y:P).Q}{\trm{y,z=x \din t}:N}
        {\Psi \mmid \Gamma,\trm{y}:P,\trm{z}:Q}{\trm{t}:N\msub{(y,z)/x}} \quad
      \fseqruule{}{\Gamma}{\trm{(d,e)}:[\Sigma(x:P). Q]}
        {\Gamma}{\trm{d}:[P] \quad}{\Gamma}{\trm{e}:[Q\msub{d/x}]} &
      \fseqrule{}{\Gamma}{\trm{\inl d}:[P \,\lor\, Q]}
        {\Gamma}{\trm{d}:[P]} \\
      \\
      \gseqruule{}{\Psi \mmid \Gamma,\trm{x}:P \,\lor\, Q}
        {\trm{\spt{x}{y.t}{z.u}}:N}
        {\Psi \mmid \Gamma,\trm{y}:P}{\trm{t}:N\msub{\inl y/x} \quad}
        {\Psi \mmid \Gamma,\trm{z}:Q}{\trm{u}:N\msub{\inr z/x}} &
      \fseqrule{}{\Gamma}{\trm{\inr d}:[P \,\lor\, Q]}
        {\Gamma}{\trm{d}:[Q]} \\
    \end{array} \vspace*{0.5em} \\
    \dotfill \\[-0.3em]
    \\
    \iruule{}{\gseq{\Psi \mmid \Gamma,\Delta\msub{d/x}}{\trm{x=d \din t}:B\msub{d/x}}}
      {\fseq{\Psi}{\trm{d}:[A]} \quad}
      {\gseq{\Psi \mmid \Gamma,\trm{x}:A,\Delta}{\trm{t}:B}}
    \qquad\quad
    \iruule{}{\gseq{\Psi \mmid \Gamma}{\trm{t\;k}:B}}
      {\gseq{\Psi \mmid \Gamma}{\trm{t}:A} \quad}
      {\fseq{\Psi,[A]}{\trm{k}:B}}\\
  \end{array} \\
  \\
  \hline
\end{array}$}
\caption{Typing rules for a dependent $\lambda$-calculus based on $\lbar$}
\label{figdeplam}
\end{figure}

We continue our investigation by adapting our variant of \LJF to dependent
types, but this unveils some issues that we will now discuss. On problem
we immediately encounter is the adaptation of the pattern machinery to the
dependent setting, mostly due to the substitutions involved in the types,
where patterns should have appeared. For the dependent implication $\Pi(x:P).N$,
using a pattern $p$ rather than a binding variable $x$ yields the question of
substituting a data structure $d$ for $p$: this becomes a much more complicated
operation than the traditional substitution. Moreover, keeping the patterns
and variables synchronised between their use in terms and in types is a
challenging task, that would probably require heavy syntactic mechanisms.
For this reason, the system shown above in Figure \ref{figdeplam} has no patterns,
but rather falls back to the traditional style of typing using only variables to
label assumptions. The language used in this variant can still be related to the
equational approach to functional programming, but the translation between
equations and terms is more involved.

The generalisation of the implication into the dependent product
$\Pi(x:P).N$ is a straightforward operation, and the rules we use are
essentially the ones found in \cite{dyckhoff:lengrand:mckinna:11:focpts} ---
except that it involves a data structure, corresponding to a focus on the
right-hand side of a sequent. Now, the case of $\Sigma$ is more complicated,
as it is \textit{a priori} unclear whether it should be obtained as a
generalisation of the negative conjunction $\land$ or of the positive
product $\times$ and both solutions might even be possible. But a generalisation
of the negative disjunction seems to be problematic, when it comes to the
specification of the second left rule, typing the $\prr\!\!\!\!$ operation.
Indeed, when focusing on $\Sigma(x:N).M$ we would need to plug a term of type
$N$ for $x$ in $M$, but this would require to maintain some \emph{``natural
deduction version''} of the term currently being transformed, and to plug
at adequate locations some translation between natural deduction style and
our sequent calculus syntax --- as done in \cite{dyckhoff:pinto:98:seqdep}.
This is quite unsatisfactory and will not help us build a proper understanding
of dependent types in a pure sequent calculus setting. The solution we adopt
here is to obtain $\Sigma(x:P).Q$ as a generalisation of the positive product
$\times$ and simply update the corresponding rules as shown in Figure
\ref{figdeplam}. The left rule is simple to define in this case, because
the decomposition of the $\Sigma$ in the context preserves the binding of
$y$ in the type $Q$.

There is a particularly interesting benefit to the use of the sequent calculus
to handle splitting as done in the left $\Sigma$ rule. Consider the elimination
rule in natural deduction:
$$\iruuuule{$\vee$e}
  {\gseq{\Gamma}{\texttt{match}~{[x.C]}~(t\,;\,y.u\,;\,z.v):C\msub{t/x}}}
  {\gseq{\Gamma,x:A\vee B}{C : \typ}}
  {\gseq{\Gamma}{t : A \vee B}}
  {\gseq{\Gamma,y:A}{u : C\msub{\inl\!\! y/x}}}
  {\gseq{\Gamma,z:B}{v : C\msub{\inr\! z/x}}}$$
and observe that it is necessary to be explicit about the return type, since
obtaining $C$ from $C\msub{t/x}$ is a complicated process, that \emph{reverses}
a substitution. This makes the term syntax heavy, while the problem is avoided
in the sequent calculus, where no substitution is needed in the conclusion.
Note that in Coq, the natural deduction style is used for the proof language,
but tactics are written in a style that is much closer to the sequent calculus
--- as this is the framework of choice for proof search --- so that tactics
have to perform some kind of translation between the two formalisms.

At the level of dependent types, there is a number of tricks used in the
Agda implementation that diverge from the proof-theoretical viewpoint. For
example, substitutions in types are treated in a complex way and may be
grouped together. Although some of the design choices can be justified by
a similarity to the focused sequent calculus, there is probably a number of
implementation techniques that have no proof-theoretical foundation.
Moreover, we have chosen here a particularly precise framework where
formulas are explicitly polarised, but in practice types in a programming
language should not always require these annotations: the question of the
presence of specific terms corresponding to shifts is therefore not obvious,
as it depends if some interesting programming constructs require their
presence or their absence. One can observe, for example, that in the system
proposed here, dependencies are subject to the presence of delays, because
of the contraction present in the left focus rule and of the treatment of
names in the $\kappa x.t$ operation.

The problem of generalising the equational style of programming associated
to the focused sequent calculus at the propositional level to the level of
dependent types is parametrised by a choice: using patterns seems to require
a complex tracking mechanism, but provides a relatively direct logical
representation of equations, while using simple variables leads to a translation
overhead. Notice however that one could think of an implementation based on
variables in which equations are easily obtained, since the language would
already be expressed in the style of the sequent calculus --- this is the
approach suggested by Epigram, where equations are meant to clarify the
meaning of programs but are not their internal representation. But we now
turn to the most challenging task of our whole enterprise: the accomodation
of induction in the framework of a focused sequent calculus in a form that
can be exploited to design declarative programs.

Induction can be expressed in Agda in a concise manner and enjoys the
benefits of the equational presentation. Consider for example the
following inductive scheme for natural numbers:
$$\begin{array}{l}
  \mathtt{ind_\mathbb{N}~:~P~zero~\imp~(\Pi(x:
    \mathbb{N}).~P~x~\imp~P~(suc~x))~\imp~\Pi(n:\mathbb{N}).~P~n}\\
  \mathtt{ind_\mathbb{N}~base~ih~zero~=~base} \\
  \mathtt{ind_\mathbb{N}~base~ih~(suc~n)~=~ih~n~(ind_\mathbb{N}~base~ih~n)}
\end{array}$$
where the code essentially relies on the matching of a natural
number, that can be either zero or the successor of another number.
It is not obvious to see through this program and select a particular
approach to induction that would be a good candidate for a proof-theoretical
description. The natural candidate for a representation of induction in
the sequent calculus would be the $\mu$ operator as studied in
\cite{baelde:phd} in the setting of intuitionistic logic. The unfocused
rules for this operator would be, from a purely logical viewpoint:
$$\gseqrule{}{\Gamma}{\mu a.B}{\Gamma}{B\msub{\mu a.B/a}}
  \qquad\quad
  \gseqrule{}{\Gamma,\mu a.B}{C}{B\msub{C/a}}{C}$$
but the presence of fixpoints has consequences for cut elimination,
as it prevents some cuts to be reduced. The usual technique applied
to avoid this problem is to build the cut rule into the left rule
for $\mu$ and to consider the result as cut free. This way, all the
cuts that cannot be reduced further are explicitly attached to the
blocking rule instance. However, the use of these rules in terms
of computation is not obvious to specify, in part because of the
complexity of the associated cut reduction, that involves the
creation of several other cuts and appeals to the functoriality
of the body $B$ of any fixpoint $\mu a.B$ --- ensured by a positivity
condition. In addition, these rules seem to interact poorly with
dependent types, as dealing with fixpoints will require a complex
handling of terms appearing inside types. It is unclear as of now
if fixpoints as expressed by $\mu$ --- and $\nu$ in the case of
induction --- can fit our scheme of explaining the implementation
of a language such as Agda, but at the same time there is no
obvious \emph{proof-theoretical} approach that accounts in a
straightforward way for the pervasive nature of inductive definitions
in the internal language of Agda, where they are handled by expansion
of names with the body of the definition.

%%%%%%%%%%%%%%%%%%%%%%%%%%%%%%%%%%%%
%%%%%%%%%%%% CONCLUSION %%%%%%%%%%%%
%%%%%%%%%%%%%%%%%%%%%%%%%%%%%%%%%%%%

\section{Conclusion and Future Work}
\label{sec:conc}

As we have seen here, the $\lbar$-calculus proposed by Herbelin as an
interpretation of the \LJT focused sequent calculus can be extended beyond
its original scope to include positive connectives, leading to a full-fledged
intuitionistic system where we can focus on the right-hand side of sequents
to decompose positives. The language we obtain is well-suited to represent
programs written in the kind of equational style found in Haskell or Agda,
the relation to equations can be made even tighter by using patterns as
labels for assumptions in the type system. The opens up the possibility to
select focused sequent calculus as a logical framework of choice for the
implementation of such languages --- as evidenced by the current state of
the implementation of Agda, containing many elements that can be explained
as sequent calculus constructs. The benefit could not only be a simplication
of such an implication, but possibly an improvement in terms of efficiency
if advanced techniques from proof theory are transferred and made practical.
Moreover, one of the strength of the logical approach is that generalisations
and extensions of all kinds are usually made simpler by the strong principles
at work: any kind of progress made on the side of proof theory could translate
into more expressive languages using the clear equational style of Haskell
and Agda --- that could be modalities, linearity or many other elements studied
in the field of computational logic.

The generalisation of this idea to handle dependent types has already
been partially investigated, but some question are left unresolved as to
the specific rules used in such a system, and the possibility of making
the system more equational by exploiting patterns. But the most difficult task
at hand is the explanation of the various treatments of induction available
in language and proofs assistants in terms of the sequent calculus. As
observed previously \cite{abel:pientka:thibodeau:setzer:14:copat}, the notion
of polarity seems to be important in the understanding of this question, but
unfortunately the proper polarised handling of fixpoints in proof theory has
yet to be found --- a number of choices are left open when it comes to the
definition of a focused system using fixpoints \cite{baelde:12:mumall}.
Note that our enterprise also yields the question of the treatment of
the identity type in proof theory, as it makes dependent pattern matching
admit the axiom $\rnm{K}$. This axiom is undesirable in homotopy type theory,
and thus the restriction of dependent pattern matching has been studied
\cite{cockx:devriese:piessens:14:nok}. But this was achieved by restricting
unification in the splitting rules, and as Agda has no explicit calculus for
splitting, this was somewhat hidden. The framework we want to develop
provides a calculus and could thus help making this restriction simpler.

\vspace*{0.3em}
{\bf Acknowledgements}. This work was funded by the grant number 10-092309
from the Danish Council for Strategic Research to the \emph{Demtech} project.
\vspace*{-0.6em}

\begin{raggedright}
\bibliographystyle{eptcs/eptcs.bst}
\bibliography{cite}
\end{raggedright}

\end{document}